%% file: jpsipolar-alicepreprint_191211.tex
\documentclass[ALICE,manyauthors]{cernphprep}
\usepackage{hyperref}

\usepackage{multirow}

\begin{document}%
%
%
\begin{titlepage}
\PHnumber{2011-182}      
\PHdate{20 December 2011}              
%
%
\title{J/$\mathbf{\psi}$ polarization
in pp collisions at $\mathbf{\sqrt{s}}$=7 TeV}
\ShortTitle{J/$\mathbf{\psi}$ polarization
in pp collisions at $\mathbf{\sqrt{s}}$=7 TeV}   
%
\Collaboration{ALICE Collaboration%
         \thanks{See Appendix~\ref{app:collab} for the list of collaboration 
                      members}}
\ShortAuthor{ALICE Collaboration}      
\begin{abstract}
We have studied ${\rm J}/\psi$ production in pp collisions at $\sqrt{s}=7$ TeV at the LHC through its muon pair decay. 
The polar and azimuthal angle distributions of the decay muons were measured, 
and results on the ${\rm J}/\psi$ polarization parameters $\lambda_{\theta}$ and $\lambda_{\phi}$ were obtained. The study was performed in the kinematic region $2.5<y<4$, $2<p_{\rm t}<8$ GeV/$c$, in the helicity and Collins-Soper reference frames. In both frames, the polarization parameters are compatible with zero,
within uncertainties.

\end{abstract}
\end{titlepage}
\setcounter{page}{2}
%
\input{jpsipolar_191211-ap.tex}               
%
\newenvironment{acknowledgement}{\relax}{\relax}
\begin{acknowledgement}
\section{Acknowledgements}
\input{acknowledgements_May2011.tex}    
\end{acknowledgement}
\newpage
%
%
\appendix
\section{The ALICE Collaboration}
\label{app:collab}
\input{authors-cernphprep_261011.tex}  
\end{document}

%% file: jpsipolar_191211-ap.tex
Almost forty years after its discovery, heavy quarkonium still represents
a challenging testing ground for models~\cite{Bra11} based on Quantum Chromodynamics (QCD). Results obtained for charmonium production at the Tevatron collider in the nineties~\cite{Abe92} led theory to recognize the role of intermediate quark-antiquark color octet states in the production process, in the framework of the Non-Relativistic QCD model (NRQCD)~\cite{Bod95}. This approach brought the calculations of $p_{\rm t}$ spectra to agree rather well with the data~\cite{Kra01} ($p_{\rm t}$ is the transvere momentum, i.e. the momentum component perpendicular to the colliding beam direction). However, the same calculations were not able to reproduce satisfactorily the polarization results for the J/$\psi$ obtained by the CDF experiment at $\sqrt{s}=1.96$ TeV~\cite{Abu07}. In particular NRQCD at leading order (LO) predicts for high-$p_{\rm t}$ J/$\psi$ ($p_{\rm t}\gg m_{\rm J/\psi}$) a significant transverse polarization, i.e. a dominant angular momentum component $J_z=\pm 1$, the $z$-axis being defined by the J/$\psi$ own momentum direction in the center of mass frame of the pp (p$\overline{\rm p}$) collision. Contrary to this expectation, CDF data~\cite{Abu07} rather exhibit a mild longitudinal polarization ($J_z= 0$).
In a recent renaissance of quarkonium studies, also related to the publication of results from RHIC at $\sqrt{s}=0.2$ TeV~\cite{Ada07}, next-to-leading order (NLO) corrections for both color singlet and color octet intermediate states were calculated, and their impact on the $p_{\rm t}$ spectra was found to be quite important~\cite{Cam07,But11,Gon09}. The influence of these corrections on the polarization calculations is expected to be significant~\cite{Gon08,Lan11} and still has not been completely worked out.
The start-up of the LHC provides the possibility to perform charmonium measurements in a new energy domain, over large ranges in $p_{\rm t}$ and rapidity ($y=0.5\ln[(E+p_{\rm z})/(E-p_{\rm z})]$, where $E$ is the energy and $p_{\rm z}$ is the momentum component parallel to the colliding beam direction). Various theoretical approaches~\cite{But11,Lan09,Vog10} proved to be rather successful in describing the first LHC experimental results on the J/$\psi$ $p_{\rm t}$ spectra~\cite{Kha11,Aai11,Aad11,Aam11}. The measurement of polarization clearly represents a more stringent test of the theoretical calculations, offering therefore the possibility of
confirming/ruling out the current QCD approach to charmonium production.

In this Letter we present the results of a study of J/$\psi$ polarization at the LHC, carried out by the ALICE experiment in pp collisions at $\sqrt{s}=7$ TeV. The ALICE experiment~\cite{Aam08} is based on a central barrel, covering the pseudorapidity region $|\eta|<0.9$~\cite{Eta} and a muon spectrometer, with $2.5<\eta <4$ coverage.
The polarization results presented in this Letter refer to inclusive J/$\psi$, measured via the ${\rm J}/\psi\rightarrow\mu^+\mu^-$ decay in the muon spectrometer. The spectrometer~\cite{Aam11} consists of a 10 interaction length 
($\lambda_{\rm I}$) thick front absorber,
to remove hadrons, followed by a 3 T$\cdot$m dipole magnet. Charged particles which exit the front absorber are tracked in a detector system made up of five stations, each one with two planes of Cathode Pad Chambers. The tracking system is followed by a 7.2~$\lambda_{\rm I}$ iron wall, which absorbs secondary hadrons escaping the front absorber and low-momentum
muons. Finally, a trigger system, based on Resistive Plate Chambers, 
is used to select candidate muons with a transverse momentum larger than a given programmable threshold. 

The analysis presented in this Letter was carried out on a significant fraction  of the 2010 sample of muon-triggered events, corresponding to an integrated luminosity $L_{\rm int}\sim 100$~nb$^{-1}$. 
The usual event selection cuts, already applied to previous analysis of J/$\psi$ production~\cite{Aam11}, were also used for the polarization study. Events with at least one vertex reconstructed in the Inner Tracker System (ITS)~\cite{Aam10}  are retained for the following analysis if they contain at least two tracks reconstructed in the muon spectrometer, out of which at least one has 
to satisfy the trigger condition (1 GeV/$c$ $p_{\rm t}$ threshold). We note that with this requirement the acceptance of the spectrometer for J/$\psi$ extends down to $p_{\rm t}=0$.
The tracks must satisfy the condition $2.5<\eta<4$ and 
must also have $17.6<R_{\rm abs}<88.9$ cm, where $R_{\rm abs}$ is the radial distance of the track from 
the beam axis at the exit of the front absorber ($z=503$ cm). The latter requirement eliminates forward tracks which, due to the high-Z material used in the absorber in that region, are strongly affected by multiple scattering. Finally, a rapidity cut $2.5<y<4$ is applied to the selected muon pairs. 

The distribution of the J/$\psi$ decay products can be expressed in its 
general form~\cite{Fac10} as
\begin{eqnarray}
W(\theta,\phi ) &\propto &\frac{1}{3+\lambda_{\theta}} \left( 1 + \lambda_{\theta}\cos^2{\theta} + \lambda_{\phi}\sin^2{\theta} \cos {2\phi}\right. \nonumber \\
&&\left. + \lambda_{\theta\phi} \sin {2\theta}\cos{\phi} \right),
\label{eq:1}
\end{eqnarray}

\noindent where $\theta$ ($\phi$) are the polar (azimuthal) angles in a given reference frame. In this analysis, the Collins-Soper (CS) and helicity (HE) frames were considered. In the CS frame the $z$-axis is defined as the bisector of the angle between the direction of one beam and the opposite of the direction of the other one, in the rest frame of the decaying particle. In the HE reference frame the $z$-axis is given by the direction of the decaying particle in the center of mass frame of the collision. The $\phi=0$ 
plane is the one containing the two beams, in the J/$\psi$ rest frame. Equation~\ref{eq:1} contains the three parameters $\lambda_{\theta}$, $\lambda_{\phi}$ and 
$\lambda_{\theta\phi}$, which quantify the degree of polarization. In particular, $\lambda_{\theta}>0$ values indicate transverse polarization, while a longitudinal polarization gives $\lambda_{\theta}<0$. In principle the values of the parameters could be extracted by means of a fit to the acceptance-corrected 2-D distributions for $\cos\theta$ vs $\phi$. However, the limited J/$\psi$ statistics (about $6.8\cdot 10^3$ signal events in the 
$p_{\rm t}$ range under study) makes a 2-D binning impossible. Therefore the study of the angular distributions was separately performed on the polar and azimuthal variables. In particular, $\lambda_{\theta}$ and $\lambda_{\phi}$ were obtained by studying the distributions 
\begin{eqnarray}
W(\cos\theta) \propto & \dfrac{1}{3+\lambda_{\theta}}\left(1+\lambda_{\theta}\cos^2{\theta}\right) \nonumber\\
W(\phi) \propto & 1+\dfrac{2\lambda_{\phi}}{3+\lambda_{\theta}}\cos{2\phi},
\label{eq:2}
\end{eqnarray}

\noindent obtained by integrating Eq.~\ref{eq:1} in the $\phi$ and $\cos{\theta}$ variables, respectively. 

The distributions of the angular variables for the J/$\psi$ decay products were  obtained starting from the study of the dimuon invariant mass spectra. The study was performed in five bins for the $|\cos\theta|$ variable (the angular distribution is symmetric with respect to $\cos\theta=0$), in the range $0<|\cos\theta|<0.8$. For the azimuthal variable four bins in $|\phi|$ were  defined, in the range $0<|\phi|<\pi/2$ (values between $\pi/2$ and $\pi$ were mirrored around $|\phi|=\pi/2$, due to the period of the $\cos 2\phi$ function). The analysis was carried out in three transverse momentum intervals ($2<p_{\rm t}<3$ GeV/$c$, $3<p_{\rm t}<4$ GeV/$c$ and $4<p_{\rm t}<8$ GeV/$c$). The limits of the explored $p_{\rm t}$ range are related to the strong decrease of the acceptance for large $|\cos\theta|$ 
values at low $p_{\rm t}$ and to the limited statistics at high $p_{\rm t}$.

\begin{figure}[htbp]
\centering
\resizebox{0.6\textwidth}{!}
{\includegraphics*[bb=10 4 567 335]{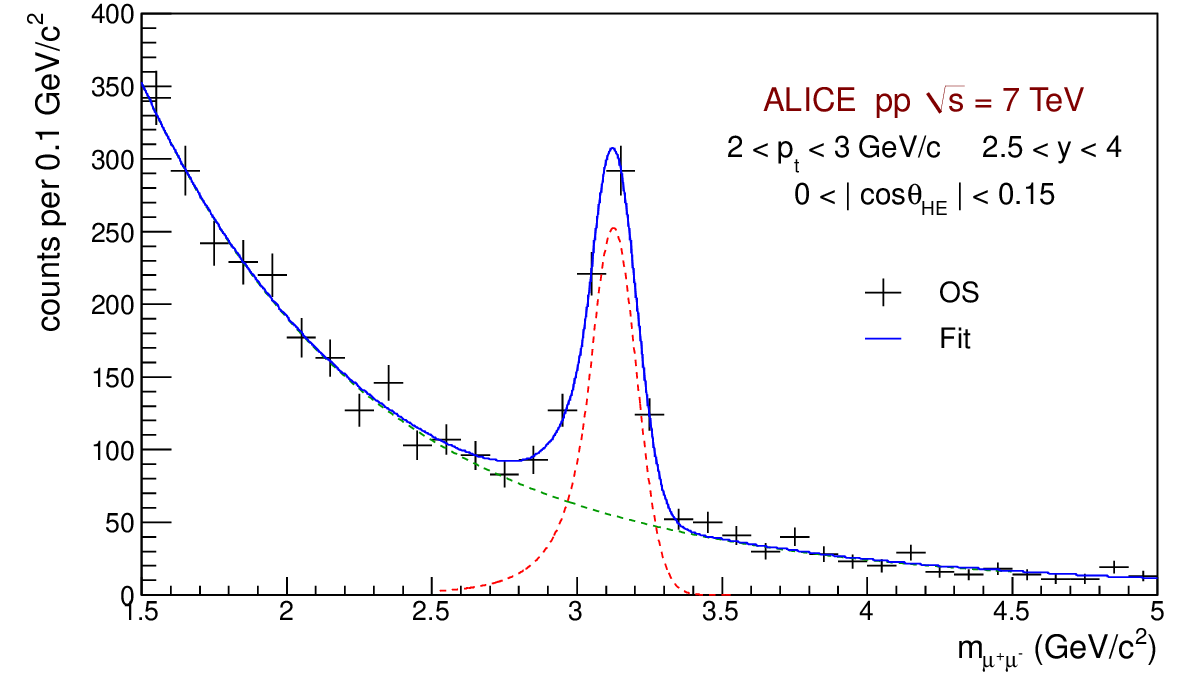}}
\caption{The dimuon invariant mass spectrum for $2<p_{\rm t}<3$ GeV/$c$, 
$0<|\cos\theta_{\rm HE}|<0.15$, together with the result of the fit. The contributions of the signal and background are also shown as dashed lines.}
\label{fig:1}
\end{figure}

The number of J/$\psi$ signal events for the various bins in $|\cos\theta|$ and $|\phi|$ were obtained by means of fits to the corresponding dimuon invariant mass spectra performed in the range $1.5<m_{\mu\mu}<5$ GeV/$c^2$, and in Fig.~\ref{fig:1} we show one of them as an example. 
The J/$\psi$ signal was described by a Crystal Ball function (CB)~\cite{Gai82}  while for the background an empirical function, corresponding to a Gaussian 
with a width linearly depending on mass, was adopted. The position of the CB peak was left as a free parameter in the fits, and was found to correspond to the nominal J/$\psi$ pole mass within at most 1\%. The width of the CB function obtained from the data (between 72 and 120 MeV/$c^2$, depending on the kinematics) was found to be in agreement with the Monte Carlo (MC) within $\sim 8-10$ MeV/$c^2$. In the fits, the width of the CB function for each bin $i$ (where $i$ represents a certain $|\cos\theta|$ or $|\phi|$ interval for the J/$\psi$ $p_{\rm t}$ bin under study) 
was fixed to $\sigma_{\rm J/\psi}^i = \sigma_{\rm J/\psi}\cdot(\sigma_{\rm J/\psi}^{i,MC}/\sigma_{\rm J/\psi}^{MC})$, i.e. by scaling the measured width for the angle-integrated spectrum with the MC ratio between the widths for the bin $i$ and for the integrated spectrum. The quality of all the 
fits is satisfactory, with $\chi^2$/ndf in a range between 0.63 and 1.34. 
Signal over background ratios in a $\pm 3\sigma$ mass window around the CB peak 
vary between 0.5 and 3.5. The number of signal events per bin ranges from $\sim$100 
(for $2<p_{\rm t}<3$ GeV/$c$, $0.6<|\cos\theta_{\rm CS}|<0.8$) to $\sim$1000 
(for $2<p_{\rm t}<3$ GeV/$c$, $0<|\cos\theta_{\rm CS}|<0.15$).

The polarization parameters for the J/$\psi$ were obtained by correcting the number of signal events $N^i_{\rm J/\psi}$ for each bin for the 
product $A_{i}\times\epsilon_{i}$ of acceptance times detection efficiency, calculated via MC simulation, and then fitting the corrected angular distributions with the functions shown in Eq.~\ref{eq:2}. The  simulation includes, for the tracking chambers, a map of dead channels and the residual misalignment of the detection elements and, for the trigger chambers, an evaluation of their efficiency based on data. It also includes a random misalignment of the tracking detector elements, of the same size of the resolution obtained by the offline alignment procedure~\cite{Aam11}. For both tracking and triggering detectors, the time variation of the efficiencies during the data taking period was accounted for (see~\cite{Aam11} for details). Since the $\cos\theta$- and $\phi$-acceptances are strongly correlated, the acceptance values as a function of one variable strongly depend on the input distribution used for the other 
variable. Given the fact that the correct input distributions are not known a priori, but rather represent the outcome of the data analysis, an iterative procedure was followed in order to determine them. 
In the first iteration a flat distribution of the angular variables (equivalent to a totally unpolarized J/$\psi$ distribution) was adopted to calculate the acceptances. After correcting the signal with those acceptances, a first determination of the polarization parameters is performed, and the results are then used in a second determination of the acceptance values. The procedure is then repeated until convergence is reached, i.e. the extracted polarization parameters do not vary by more than 0.005 between two successive iterations. This occurs, for this analysis, after at most three steps. It was also checked that using polarized MC input distributions in the first iteration the procedure converges towards the same results as in the default, unpolarized, case.
Typical $A_{i}\times\epsilon_{i}$ values vary between $\sim$0.22 (0.05) at low $p_{\rm t}$ and large $|\cos\theta|$ and
$\sim$0.41 (0.63) at large $p_{\rm t}$ and small $|\cos\theta|$ for the HE (CS) frame. 

\begin{figure}[htbp]
\centering
\resizebox{0.7\textwidth}{!}
{\includegraphics*[bb=4 19 588 420]{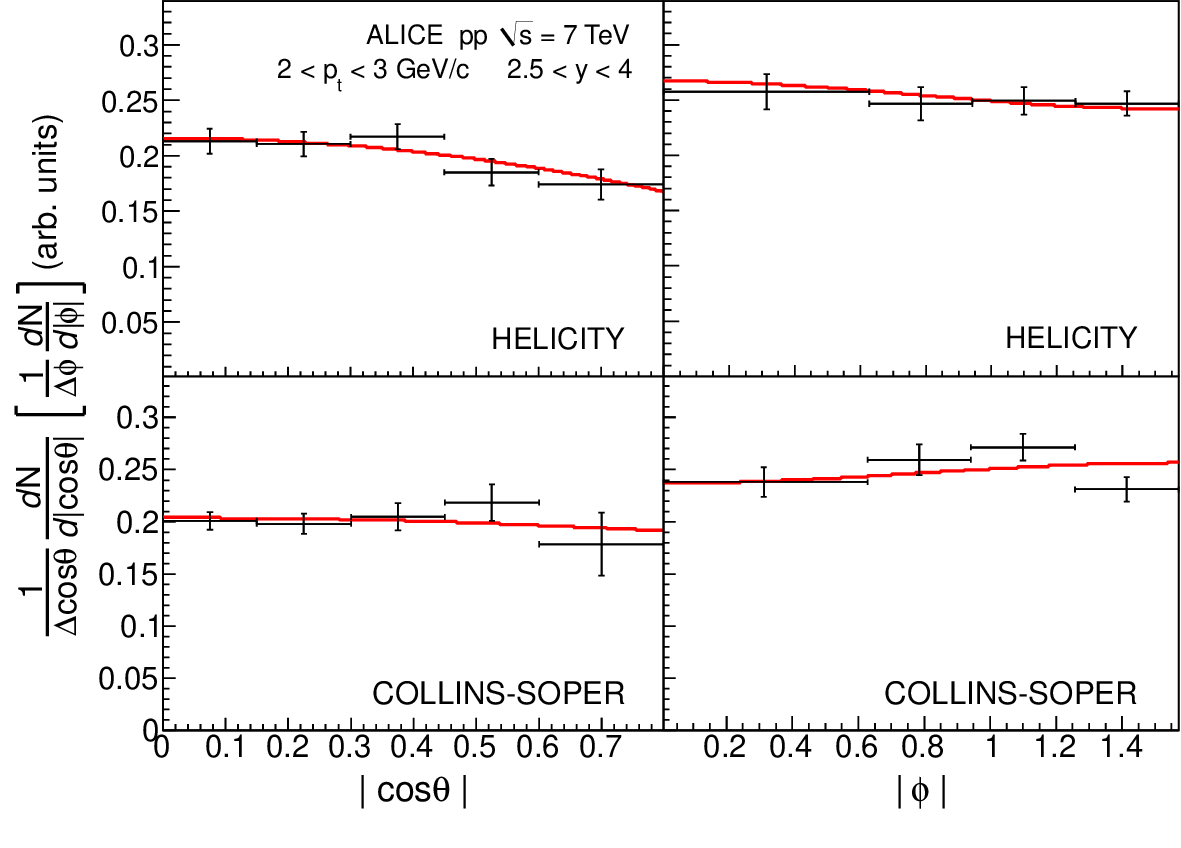}}
\caption{The acceptance corrected angular distributions for the J/$\psi$ decay muons, for $2<p_{\rm t}<3$ GeV/$c$. The simultaneous fit to the results in the CS and HE frames is also shown. The plotted errors are purely statistical. The horizontal bars represent the bin width.}
\label{fig:2}
\end{figure}

A simultaneous study of the J/$\psi$ polarization variables in several reference frames, as first carried out in hadroproduction studies by the HERA-B experiment~\cite{Abt09}, is particularly interesting since 
consistency checks on the results can be performed, using combinations of the polarization parameters which are frame-invariant. 
In particular we made use of the invariant $F=(\lambda_{\theta}+3\lambda_{\phi})/(1-\lambda_{\phi})$~\cite{Fac10}, performing a simultaneous fit of the $|\cos{\theta}|$ and $|\phi|$ distributions in the two reference systems and further constraining the fit by imposing $F$ to be the same in the CS and HE frames. In Fig.~\ref{fig:2} we present, as an example, the result of 
such a fit relative to the last iteration of the $A_{i}\times\epsilon_{i}$ calculation, for $2<p_{\rm t}<3$ GeV/$c$. The $\chi^2/ndf$ values ($ndf=10$) are 1.08, 1.00, 1.32 for $2<p_{\rm t}<3$, $3<p_{\rm t}<4$ and $4<p_{\rm t}<8$ GeV/$c$, respectively, showing that the quality of the fits is good. Compatible results are obtained when the constraint on $F$ is released.

In the analysis described so far, the $\lambda_{\theta\phi}$ parameter was  implicitly assumed to be zero in the iterative acceptance calculation. In the one-dimensional approach followed in this analysis $\lambda_{\theta\phi}$ could be estimated from the data, defining an ad-hoc variable $\tilde{\phi}$ which is a function of $\cos\theta$ and $\phi$ and contains $\lambda_{\theta\phi}$ as a parameter (see~\cite{Fac10} for details).
In principle, the iterative procedure applied to $\lambda_{\theta}$ and $\lambda_{\phi}$ determination could be extended to include $\lambda_{\theta\phi}$; however, in some cases, relatively small statistical fluctuations in the distributions of the measured variables tend to induce 
large variations of the fitted values in the following iterations, leading to convergence problems. A check of the $\lambda_{\theta\phi}=0$ assumption was done a posteriori for each $p_{\rm t}$ bin, by fitting the $\tilde{\phi}$ distributions, corrected with an acceptance which makes use of the measured $\lambda_{\theta}$ and $\lambda_{\phi}$ values as inputs. In this way we get 
for all the $p_{\rm t}$ bins $\lambda_{\theta\phi}$ values compatible with zero for both CS and HE reference frames. We also note that all the previous experiments assumed $\lambda_{\theta\phi}=0$ in their analysis, with the exception of HERA-B~\cite{Abt09} who measured it in pA collisions at $\sqrt{s}=41.6$ GeV and found values ranging from 0 to 0.05.

Various sources of systematic uncertainty on the measurement of the polarization parameters have been investigated. 
The uncertainty on the signal extraction was studied by leaving in the fits the width of the CB function as a free parameter. This choice leads to an absolute variation of the polarization parameters between 0.02 and 0.10. 
Another sizeable source of systematic uncertainty is the choice of the input distributions for $p_{\rm t}$ 
and $y$ in the simulation. It was evaluated by comparing the results obtained with a parameterization of our 7 TeV results on differential J/$\psi$ cross sections~\cite{Aam11} with those obtained using an extrapolation of lower 
energy results~\cite{Bos11}. The absolute effect on the polarization parameters varies between 0.01 and 0.07.
For the lowest $p_{\rm t}$ bin, the acceptance in the HE frame drops by about 40\% 
in the highest $|\cos\theta|$ bin used in the analysis ($0.6<|\cos\theta|<0.8$), and has also a strong variation inside the bin itself. We therefore followed an alternative approach, fitting the angular spectrum in the 
restricted interval $0<|\cos\theta|<0.6$ (instead of the default choice $0<|\cos\theta|<0.8$) and we conservatively considered the variation in 
the result of the fit (0.15) as an additional systematic uncertainty on $\lambda_{\theta}$. For consistency, the same evaluation was performed in the CS frame. 
The role of the systematic uncertainties on the trigger and tracking 
efficiency~\cite{Aam11} was also studied. The first was evaluated by varying the efficiency values for each detector element by 2\% with respect to the default values in the simulation. This choice is related to the estimated uncertainty on the detector efficiency calculation.
For the second we have used the rather conservative choice of comparing the reference results, 
obtained with realistic dead channel maps, with those relative to an ideal detector set-up. The result is typically 0.03-0.04. Finally, by quadratically combining the results for the various sources, values between 0.04 and 0.21 are obtained for the global systematic uncertainties.

\begin{figure}[htbp]
\vskip -2cm
\centering
\resizebox{0.55\textwidth}{!}
{\includegraphics*[bb=10 25 544 667]{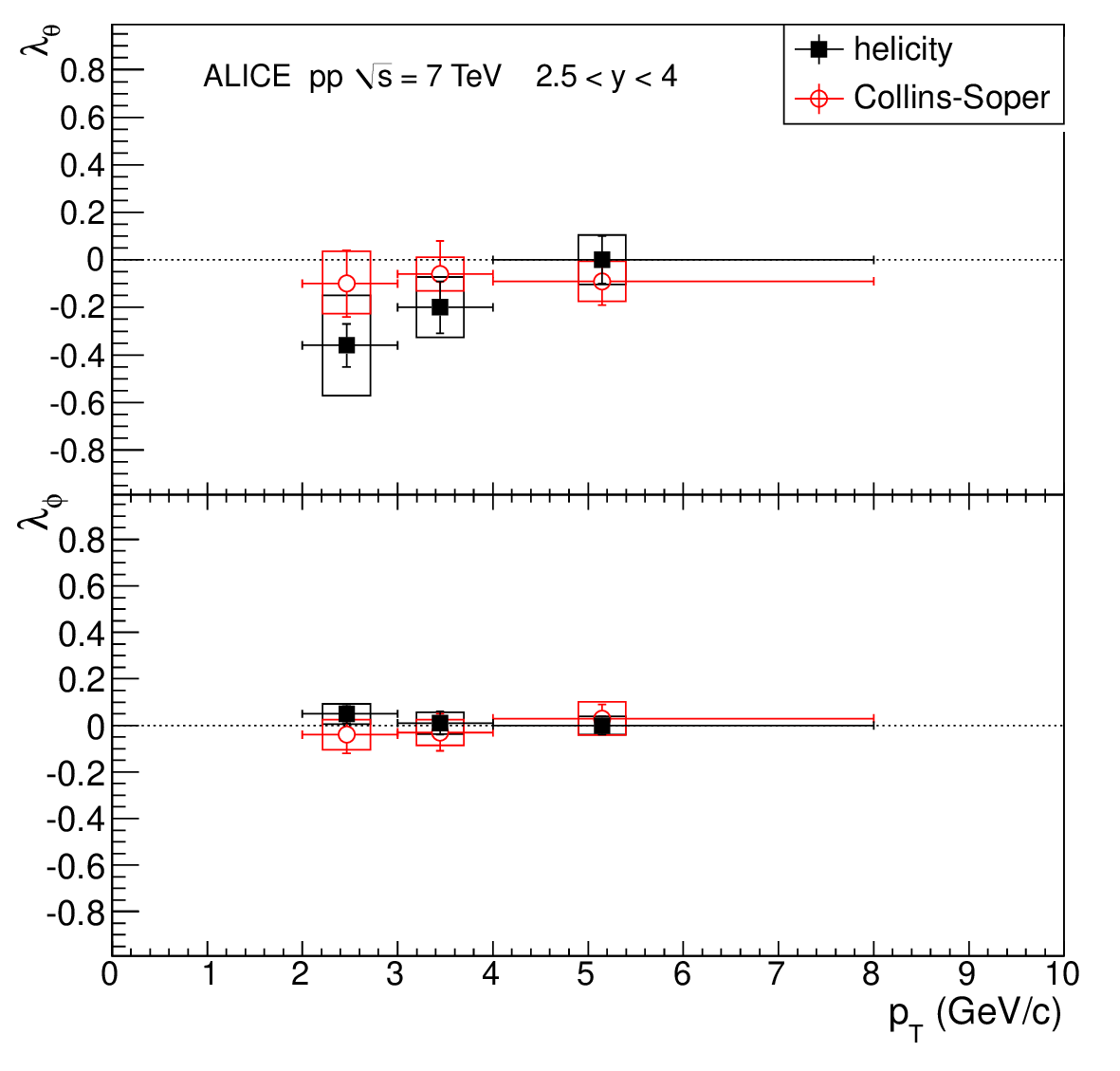}}
\caption{$\lambda_{\theta}$ and $\lambda_{\phi}$ as a function of $p_{\rm t}$ for inclusive J/$\psi$, measured in the HE (closed squares) and CS (open circles) frames. The error bars represent statistical errors, while systematic uncertainties are shown as boxes.}
\label{fig:3}
\end{figure}

In Fig.~\ref{fig:3} we show the results on $\lambda_{\theta}$ and $\lambda_{\phi}$ for inclusive J/$\psi$ production. 
In both frames  
all the parameters are compatible with zero, with a possible hint for a longitudinal polarization at low $p_{\rm t}$ (at a 1.6$\sigma$ level) in the HE frame. The numerical values are given in Table~\ref{tab:2}.

\begin{table}[htbp]
\centering
\caption{\label{tab:2} The values of $\lambda_{\theta}$ and $\lambda_{\phi}$ in the two reference frames. Statistical and systematic uncertainties are quoted separately.}
\begin{tabular}{cccc}
& $p_{\rm t}$ ($\langle p_{\rm t}\rangle$) & \multirow{2}{*}{$\lambda_{\theta}$} & \multirow{2}{*}{$\lambda_{\phi}$} \\
& (GeV/$c$)             & &  \\
\hline
& 2-3 (2.5) & $-0.36\pm 0.09\pm 0.21$ & $0.05\pm 0.04\pm 0.04$ \\
HE & 3-4 (3.4) & $-0.20\pm 0.11 \pm 0.13$ & $0.01\pm 0.05\pm 0.05$ \\
& 4-8 (5.1) & $0.00\pm 0.10\pm 0.10$ & $0.00\pm 0.04\pm 0.04$ \\
\hline
& 2-3 (2.5) & $-0.10\pm 0.14\pm 0.13$ & $-0.04\pm 0.08\pm 0.07$ \\
CS & 3-4 (3.4) & $-0.06\pm 0.14 \pm 0.07$ & $-0.03\pm 0.08\pm 0.05$ \\
& 4-8 (5.1)& $-0.09\pm 0.10\pm 0.08$ & $0.03\pm 0.06\pm 0.07$ \\
\end{tabular}
\end{table}

The inclusive J/$\psi$ yield is 
composed of a ``prompt'' component (direct J/$\psi$ $+$ decay of the $\psi(2S)$ and $\chi_c$ resonances) and of a component from B-meson decays. 
In the $p_{\rm t}$ range accessed in this analysis the B-meson decay component accounts for 10\% ($2<p_{\rm T}<3$ GeV/$c$), 12\% ($3<p_{\rm T}<4$ GeV/$c$) and 15\% ($4<p_{\rm T}<8$ GeV/$c$) of the inclusive yield, according to the LHCb measurements carried out in our same kinematical domain~\cite{Aai11}. The polarization of the non-prompt component is expected to be quite small. In fact, even if a sizeable polarization were observed when the polarization axis refers to the B-meson direction~\cite{Aub03}, it would be strongly smeared when it is calculated with respect to the direction of the decay J/$\psi$~\cite{Aai11}, as observed by CDF, who measured in this way $\lambda_{\theta}(J/\psi\leftarrow B)\sim -0.1$ in the HE frame~\cite{Abu07}.
Assuming conservatively $|\lambda_{\theta}(J/\psi\leftarrow B)|<0.2$ for both frames, and taking into account the fraction of the inclusive yield coming from B-meson decays~\cite{Aai11}, the difference between prompt and inclusive J/$\psi$ polarization was estimated and found to be at most 0.05, a value smaller than the systematic uncertainties of our measurements.
Concerning higher-mass charmonia, the $\chi_c\rightarrow {\rm J/\psi}+\gamma$ decay cannot be reconstructed in the muon spectrometer, and the $\psi(2S)\rightarrow\mu\mu$ statistics is currently too low. Values of the feed-down ratios measured mainly by lower energy experiments range from $\sim$10\% for the $\psi(2S)$~\cite{Aal09} to 25-30\% for the $\chi_c$~\cite{Fac08}, implying that there could be a sizeable difference between direct and prompt J/$\psi$ polarization.

The results presented in Fig.~\ref{fig:3} extend the study of the J/$\psi$ polarization to LHC energies and therefore open up a new testing ground for theoretical models. 
At present, NLO calculations for direct J/$\psi$ polarization at the LHC via the color-singlet channel~\cite{Lan09,Gon08} predict a large longitudinal polarization in the HE frame ($\lambda_{\theta}\sim -0.6$) at $p_{\rm t}\sim$ 5  GeV/$c$, which is in contrast with the vanishing polarization that we observe in such a transverse momentum region. 
The contribution of the S-wave color-octet channels was also worked out~\cite{Gon09} and indicates a significantly different trend (large transverse polarization) with respect to the color-singlet contribution, but again in contrast with our result.
In this situation, a rigorous treatment on the theory side of all the color-octet terms (including P-wave contributions) is mandatory, as well as a study of the contribution of $\chi_c$ and $\psi(2S)$ feed-down which, as outlined before, is important for a quantitative comparison with our result~\cite{ButPC}. 
Such studies are presently in progress and the comparison of their outcome with the results presented in this Letter will allow a very significant test of the 
understanding of the heavy-quarkonium production mechanisms in QCD-based models.

In summary, we have measured the polarization parameters $\lambda_{\theta}$ and $\lambda_{\phi}$ for inclusive J/$\psi$ production in $\sqrt{s}=7$ TeV pp collisions at the LHC. The measurement was carried out in the kinematical region $2.5<y<4$, $2<p_{\rm t}<8$ GeV/$c$. The polarization parameters $\lambda_{\theta}$ and $\lambda_{\phi}$ are consistent with zero, in both the helicity and Collins-Soper reference frames. These results can be used as a stringent constraint on the commonly adopted QCD framework for heavy quarkonium production.

%% file: acknowledgements_May2011.tex
The ALICE Collaboration would like to thank all its engineers and technicians for their invaluable contributions to the construction of the experiment and the CERN accelerator teams for the outstanding performance of the LHC complex.
\\
The ALICE Collaboration acknowledges the following funding agencies for their support in building and
running the ALICE detector:
 \\
Department of Science and Technology, South Africa;
 \\
Calouste Gulbenkian Foundation from Lisbon and Swiss Fonds Kidagan, Armenia;
 \\
Conselho Nacional de Desenvolvimento Cient\'{\i}fico e Tecnol\'{o}gico (CNPq), Financiadora de Estudos e Projetos (FINEP),
Funda\c{c}\~{a}o de Amparo \`{a} Pesquisa do Estado de S\~{a}o Paulo (FAPESP);
 \\
National Natural Science Foundation of China (NSFC), the Chinese Ministry of Education (CMOE)
and the Ministry of Science and Technology of China (MSTC);
 \\
Ministry of Education and Youth of the Czech Republic;
 \\
Danish Natural Science Research Council, the Carlsberg Foundation and the Danish National Research Foundation;
 \\
The European Research Council under the European Community's Seventh Framework Programme;
 \\
Helsinki Institute of Physics and the Academy of Finland;
 \\
French CNRS-IN2P3, the `Region Pays de Loire', `Region Alsace', `Region Auvergne' and CEA, France;
 \\
German BMBF and the Helmholtz Association;
\\
General Secretariat for Research and Technology, Ministry of
Development, Greece;
\\
Hungarian OTKA and National Office for Research and Technology (NKTH);
 \\
Department of Atomic Energy and Department of Science and Technology of the Government of India;
 \\
Istituto Nazionale di Fisica Nucleare (INFN) of Italy;
 \\
MEXT Grant-in-Aid for Specially Promoted Research, Ja\-pan;
 \\
Joint Institute for Nuclear Research, Dubna;
 \\
National Research Foundation of Korea (NRF);
 \\
CONACYT, DGAPA, M\'{e}xico, ALFA-EC and the HELEN Program (High-Energy physics Latin-American--European Network);
 \\
Stichting voor Fundamenteel Onderzoek der Materie (FOM) and the Nederlandse Organisatie voor Wetenschappelijk Onderzoek (NWO), Netherlands;
 \\
Research Council of Norway (NFR);
 \\
Polish Ministry of Science and Higher Education;
 \\
National Authority for Scientific Research - NASR (Autoritatea Na\c{t}ional\u{a} pentru Cercetare \c{S}tiin\c{t}ific\u{a} - ANCS);
 \\
Federal Agency of Science of the Ministry of Education and Science of Russian Federation, International Science and
Technology Center, Russian Academy of Sciences, Russian Federal Agency of Atomic Energy, Russian Federal Agency for Science and Innovations and CERN-INTAS;
 \\
Ministry of Education of Slovakia;
 \\
CIEMAT, EELA, Ministerio de Educaci\'{o}n y Ciencia of Spain, Xunta de Galicia (Conseller\'{\i}a de Educaci\'{o}n),
CEA\-DEN, Cubaenerg\'{\i}a, Cuba, and IAEA (International Atomic Energy Agency);
 \\
Swedish Research Council (VR) and Knut $\&$ Alice Wallenberg Foundation (KAW);
 \\
Ukraine Ministry of Education and Science;
 \\
United Kingdom Science and Technology Facilities Council (STFC);
 \\
The United States Department of Energy, the United States National
Science Foundation, the State of Texas, and the State of Ohio.

%% file: authors-cernphprep_261011.tex
\begingroup
\small
\begin{flushleft}
B.~Abelev\Irefn{org1234}\And
A.~Abrahantes~Quintana\Irefn{org1197}\And
D.~Adamov\'{a}\Irefn{org1283}\And
A.M.~Adare\Irefn{org1260}\And
M.M.~Aggarwal\Irefn{org1157}\And
G.~Aglieri~Rinella\Irefn{org1192}\And
A.G.~Agocs\Irefn{org1143}\And
A.~Agostinelli\Irefn{org1132}\And
S.~Aguilar~Salazar\Irefn{org1247}\And
Z.~Ahammed\Irefn{org1225}
% ...
\renewcommand\labelenumi{\textsuperscript{\theenumi}~}
\section*{Affiliation notes}
\renewcommand\theenumi{\roman{enumi}}
\begin{Authlist}
\item \Adef{0}Deceased
\item \Adef{Dipartimento di Fisica dell'Universita, Udine, Italy}Also at: Dipartimento di Fisica dell'Universita, Udine, Italy
\item \Adef{M.V.Lomonosov Moscow State University, D.V.Skobeltsyn Institute of Nuclear Physics, Moscow, Russia}Also at: M.V.Lomonosov Moscow State University, D.V.Skobeltsyn Institute of Nuclear Physics, Moscow, Russia
\item \Adef{Institute of Nuclear Sciences, Belgrade, Serbia}Also at: "Vin\v{c}a" Institute of Nuclear Sciences, Belgrade, Serbia
\end{Authlist}
\section*{Collaboration Institutes}
\renewcommand\theenumi{\arabic{enumi}~}
\begin{Authlist}
\item \Idef{org1279}Benem\'{e}rita Universidad Aut\'{o}noma de Puebla, Puebla, Mexico
\item \Idef{org1220}Bogolyubov Institute for Theoretical Physics, Kiev, Ukraine
\item \Idef{org1262}Budker Institute for Nuclear Physics, Novosibirsk, Russia
\item \Idef{org1292}California Polytechnic State University, San Luis Obispo, California, United States
\item \Idef{org14939}Centre de Calcul de l'IN2P3, Villeurbanne, France
\item \Idef{org1197}Centro de Aplicaciones Tecnol\'{o}gicas y Desarrollo Nuclear (CEADEN), Havana, Cuba
\item \Idef{org1242}Centro de Investigaciones Energ\'{e}ticas Medioambientales y Tecnol\'{o}gicas (CIEMAT), Madrid, Spain
\item \Idef{org1244}Centro de Investigaci\'{o}n y de Estudios Avanzados (CINVESTAV), Mexico City and M\'{e}rida, Mexico
\item \Idef{org1335}Centro Fermi -- Centro Studi e Ricerche e Museo Storico della Fisica ``Enrico Fermi'', Rome, Italy
\item \Idef{org17347}Chicago State University, Chicago, United States
\item \Idef{org1118}China Institute of Atomic Energy, Beijing, China
\item \Idef{org1288}Commissariat \`{a} l'Energie Atomique, IRFU, Saclay, France
\item \Idef{org1294}Departamento de F\'{\i}sica de Part\'{\i}culas and IGFAE, Universidad de Santiago de Compostela, Santiago de Compostela, Spain
\item \Idef{org1106}Department of Physics Aligarh Muslim University, Aligarh, India
\item \Idef{org1121}Department of Physics and Technology, University of Bergen, Bergen, Norway
\item \Idef{org1162}Department of Physics, Ohio State University, Columbus, Ohio, United States
\item \Idef{org1300}Department of Physics, Sejong University, Seoul, South Korea
\item \Idef{org1268}Department of Physics, University of Oslo, Oslo, Norway
\item \Idef{org1132}Dipartimento di Fisica dell'Universit\`{a} and Sezione INFN, Bologna, Italy
\item \Idef{org1315}Dipartimento di Fisica dell'Universit\`{a} and Sezione INFN, Trieste, Italy
\item \Idef{org1145}Dipartimento di Fisica dell'Universit\`{a} and Sezione INFN, Cagliari, Italy
\item \Idef{org1270}Dipartimento di Fisica dell'Universit\`{a} and Sezione INFN, Padova, Italy
\item \Idef{org1285}Dipartimento di Fisica dell'Universit\`{a} `La Sapienza' and Sezione INFN, Rome, Italy
\item \Idef{org1154}Dipartimento di Fisica e Astronomia dell'Universit\`{a} and Sezione INFN, Catania, Italy
\item \Idef{org1290}Dipartimento di Fisica `E.R.~Caianiello' dell'Universit\`{a} and Gruppo Collegato INFN, Salerno, Italy
\item \Idef{org1312}Dipartimento di Fisica Sperimentale dell'Universit\`{a} and Sezione INFN, Turin, Italy
\item \Idef{org1103}Dipartimento di Scienze e Tecnologie Avanzate dell'Universit\`{a} del Piemonte Orientale and Gruppo Collegato INFN, Alessandria, Italy
\item \Idef{org1114}Dipartimento Interateneo di Fisica `M.~Merlin' and Sezione INFN, Bari, Italy
\item \Idef{org1237}Division of Experimental High Energy Physics, University of Lund, Lund, Sweden
\item \Idef{org1192}European Organization for Nuclear Research (CERN), Geneva, Switzerland
\item \Idef{org1227}Fachhochschule K\"{o}ln, K\"{o}ln, Germany
\item \Idef{org1122}Faculty of Engineering, Bergen University College, Bergen, Norway
\item \Idef{org1136}Faculty of Mathematics, Physics and Informatics, Comenius University, Bratislava, Slovakia
\item \Idef{org1274}Faculty of Nuclear Sciences and Physical Engineering, Czech Technical University in Prague, Prague, Czech Republic
\item \Idef{org1229}Faculty of Science, P.J.~\v{S}af\'{a}rik University, Ko\v{s}ice, Slovakia
\item \Idef{org1184}Frankfurt Institute for Advanced Studies, Johann Wolfgang Goethe-Universit\"{a}t Frankfurt, Frankfurt, Germany
\item \Idef{org1215}Gangneung-Wonju National University, Gangneung, South Korea
\item \Idef{org1212}Helsinki Institute of Physics (HIP) and University of Jyv\"{a}skyl\"{a}, Jyv\"{a}skyl\"{a}, Finland
\item \Idef{org1203}Hiroshima University, Hiroshima, Japan
\item \Idef{org1329}Hua-Zhong Normal University, Wuhan, China
\item \Idef{org1254}Indian Institute of Technology, Mumbai, India
\item \Idef{org1266}Institut de Physique Nucl\'{e}aire d'Orsay (IPNO), Universit\'{e} Paris-Sud, CNRS-IN2P3, Orsay, France
\item \Idef{org1277}Institute for High Energy Physics, Protvino, Russia
\item \Idef{org1249}Institute for Nuclear Research, Academy of Sciences, Moscow, Russia
\item \Idef{org1320}Nikhef, National Institute for Subatomic Physics and Institute for Subatomic Physics of Utrecht University, Utrecht, Netherlands
\item \Idef{org1250}Institute for Theoretical and Experimental Physics, Moscow, Russia
\item \Idef{org1230}Institute of Experimental Physics, Slovak Academy of Sciences, Ko\v{s}ice, Slovakia
\item \Idef{org1127}Institute of Physics, Bhubaneswar, India
\item \Idef{org1275}Institute of Physics, Academy of Sciences of the Czech Republic, Prague, Czech Republic
\item \Idef{org1139}Institute of Space Sciences (ISS), Bucharest, Romania
\item \Idef{org27399}Institut f\"{u}r Informatik, Johann Wolfgang Goethe-Universit\"{a}t Frankfurt, Frankfurt, Germany
\item \Idef{org1185}Institut f\"{u}r Kernphysik, Johann Wolfgang Goethe-Universit\"{a}t Frankfurt, Frankfurt, Germany
\item \Idef{org1177}Institut f\"{u}r Kernphysik, Technische Universit\"{a}t Darmstadt, Darmstadt, Germany
\item \Idef{org1256}Institut f\"{u}r Kernphysik, Westf\"{a}lische Wilhelms-Universit\"{a}t M\"{u}nster, M\"{u}nster, Germany
\item \Idef{org1246}Instituto de Ciencias Nucleares, Universidad Nacional Aut\'{o}noma de M\'{e}xico, Mexico City, Mexico
\item \Idef{org1247}Instituto de F\'{\i}sica, Universidad Nacional Aut\'{o}noma de M\'{e}xico, Mexico City, Mexico
\item \Idef{org23333}Institut of Theoretical Physics, University of Wroclaw
\item \Idef{org1308}Institut Pluridisciplinaire Hubert Curien (IPHC), Universit\'{e} de Strasbourg, CNRS-IN2P3, Strasbourg, France
\item \Idef{org1182}Joint Institute for Nuclear Research (JINR), Dubna, Russia
\item \Idef{org1143}KFKI Research Institute for Particle and Nuclear Physics, Hungarian Academy of Sciences, Budapest, Hungary
\item \Idef{org1199}Kirchhoff-Institut f\"{u}r Physik, Ruprecht-Karls-Universit\"{a}t Heidelberg, Heidelberg, Germany
\item \Idef{org1160}Laboratoire de Physique Corpusculaire (LPC), Clermont Universit\'{e}, Universit\'{e} Blaise Pascal, CNRS--IN2P3, Clermont-Ferrand, France
\item \Idef{org1194}Laboratoire de Physique Subatomique et de Cosmologie (LPSC), Universit\'{e} Joseph Fourier, CNRS-IN2P3, Institut Polytechnique de Grenoble, Grenoble, France
\item \Idef{org1187}Laboratori Nazionali di Frascati, INFN, Frascati, Italy
\item \Idef{org1232}Laboratori Nazionali di Legnaro, INFN, Legnaro, Italy
\item \Idef{org1125}Lawrence Berkeley National Laboratory, Berkeley, California, United States
\item \Idef{org1234}Lawrence Livermore National Laboratory, Livermore, California, United States
\item \Idef{org1251}Moscow Engineering Physics Institute, Moscow, Russia
\item \Idef{org1140}National Institute for Physics and Nuclear Engineering, Bucharest, Romania
\item \Idef{org1165}Niels Bohr Institute, University of Copenhagen, Copenhagen, Denmark
\item \Idef{org1109}Nikhef, National Institute for Subatomic Physics, Amsterdam, Netherlands
\item \Idef{org1283}Nuclear Physics Institute, Academy of Sciences of the Czech Republic, \v{R}e\v{z} u Prahy, Czech Republic
\item \Idef{org1264}Oak Ridge National Laboratory, Oak Ridge, Tennessee, United States
\item \Idef{org1189}Petersburg Nuclear Physics Institute, Gatchina, Russia
\item \Idef{org1170}Physics Department, Creighton University, Omaha, Nebraska, United States
\item \Idef{org1157}Physics Department, Panjab University, Chandigarh, India
\item \Idef{org1112}Physics Department, University of Athens, Athens, Greece
\item \Idef{org1152}Physics Department, University of Cape Town, iThemba LABS, Cape Town, South Africa
\item \Idef{org1209}Physics Department, University of Jammu, Jammu, India
\item \Idef{org1207}Physics Department, University of Rajasthan, Jaipur, India
\item \Idef{org1200}Physikalisches Institut, Ruprecht-Karls-Universit\"{a}t Heidelberg, Heidelberg, Germany
\item \Idef{org1325}Purdue University, West Lafayette, Indiana, United States
\item \Idef{org1281}Pusan National University, Pusan, South Korea
\item \Idef{org1176}Research Division and ExtreMe Matter Institute EMMI, GSI Helmholtzzentrum f\"ur Schwerionenforschung, Darmstadt, Germany
\item \Idef{org1334}Rudjer Bo\v{s}kovi\'{c} Institute, Zagreb, Croatia
\item \Idef{org1298}Russian Federal Nuclear Center (VNIIEF), Sarov, Russia
\item \Idef{org1252}Russian Research Centre Kurchatov Institute, Moscow, Russia
\item \Idef{org1224}Saha Institute of Nuclear Physics, Kolkata, India
\item \Idef{org1130}School of Physics and Astronomy, University of Birmingham, Birmingham, United Kingdom
\item \Idef{org1338}Secci\'{o}n F\'{\i}sica, Departamento de Ciencias, Pontificia Universidad Cat\'{o}lica del Per\'{u}, Lima, Peru
\item \Idef{org1146}Sezione INFN, Cagliari, Italy
\item \Idef{org1115}Sezione INFN, Bari, Italy
\item \Idef{org1313}Sezione INFN, Turin, Italy
\item \Idef{org1133}Sezione INFN, Bologna, Italy
\item \Idef{org1155}Sezione INFN, Catania, Italy
\item \Idef{org1316}Sezione INFN, Trieste, Italy
\item \Idef{org1286}Sezione INFN, Rome, Italy
\item \Idef{org1271}Sezione INFN, Padova, Italy
\item \Idef{org1322}Soltan Institute for Nuclear Studies, Warsaw, Poland
\item \Idef{org1258}SUBATECH, Ecole des Mines de Nantes, Universit\'{e} de Nantes, CNRS-IN2P3, Nantes, France
\item \Idef{org1304}Technical University of Split FESB, Split, Croatia
\item \Idef{org1168}The Henryk Niewodniczanski Institute of Nuclear Physics, Polish Academy of Sciences, Cracow, Poland
\item \Idef{org17361}The University of Texas at Austin, Physics Department, Austin, TX, United States
\item \Idef{org1173}Universidad Aut\'{o}noma de Sinaloa, Culiac\'{a}n, Mexico
\item \Idef{org1296}Universidade de S\~{a}o Paulo (USP), S\~{a}o Paulo, Brazil
\item \Idef{org1149}Universidade Estadual de Campinas (UNICAMP), Campinas, Brazil
\item \Idef{org1239}Universit\'{e} de Lyon, Universit\'{e} Lyon 1, CNRS/IN2P3, IPN-Lyon, Villeurbanne, France
\item \Idef{org1205}University of Houston, Houston, Texas, United States
\item \Idef{org1222}University of Tennessee, Knoxville, Tennessee, United States
\item \Idef{org1310}University of Tokyo, Tokyo, Japan
\item \Idef{org1318}University of Tsukuba, Tsukuba, Japan
\item \Idef{org21360}Eberhard Karls Universit\"{a}t T\"{u}bingen, T\"{u}bingen, Germany
\item \Idef{org1225}Variable Energy Cyclotron Centre, Kolkata, India
\item \Idef{org1306}V.~Fock Institute for Physics, St. Petersburg State University, St. Petersburg, Russia
\item \Idef{org1323}Warsaw University of Technology, Warsaw, Poland
\item \Idef{org1179}Wayne State University, Detroit, Michigan, United States
\item \Idef{org1260}Yale University, New Haven, Connecticut, United States
\item \Idef{org1332}Yerevan Physics Institute, Yerevan, Armenia
\item \Idef{org15649}Yildiz Technical University, Istanbul, Turkey
\item \Idef{org1301}Yonsei University, Seoul, South Korea
\item \Idef{org1327}Zentrum f\"{u}r Technologietransfer und Telekommunikation (ZTT), Fachhochschule Worms, Worms, Germany
\end{Authlist}
\endgroup

%% file: jpsipolar-alicepreprint_191211.bbl
\begin{thebibliography}{99}
\bibitem{Bra11} N.~Brambilla et al., Eur. Phys. J. {\bf C71}(2011) 1534.
\bibitem{Abe92} F.~Abe et al. (CDF Collaboration), Phys. Rev. Lett. {\bf 69} (1992) 3704; F.~Abe et al. (CDF Collaboration), Phys. Rev. Lett. {\bf 79} (1997) 572; S.~Abachi et al. (D0 Collaboration), Phys. Lett. {\bf B370} (1996) 239.
\bibitem{Bod95} G.~Bodwin, E.~Braaten and G.P.~Lepage, Phys. Rev. {\bf D51} (1995) 1125-1171; Erratum-ibid. {\bf D55} (1997) 5853.
\bibitem{Kra01} M.~Kr\"amer, Prog. Part. Nucl. Phys. {\bf 47} (2001) 141. 
\bibitem{Abu07} A.~Abulencia et al. (CDF Collaboration), Phys. Rev. Lett. 
{\bf 99} (2007) 132001. 
\bibitem{Ada07} A.~Adare et al. (PHENIX Collaboration), Phys. Rev. Lett. {\bf 98} (2007) 232002; 
A.~Adare et al. (PHENIX Collaboration), Phys. Rev. {\bf D82} (2010) 012001.
\bibitem{Cam07} J.~Campbell, F.~Maltoni, F.~Tramontano, Phys. Rev. Lett. 
{\bf 98} (2007) 252002.
\bibitem{But11} M.~Butensch\"on, B.A.~Kniehl, Phys. Rev. Lett. {\bf 106} (2011) 022003; 
 Y.Q.~Ma, K.~Wang, K.T.~Chao, Phys. Rev. Lett. {\bf 106} (2011) 042002.
\bibitem{Gon09} B.~Gong, X.Q.~Li, J.-X.~Wang, Phys. Lett. {\bf B673} (2009) 197.
\bibitem{Gon08} B.~Gong, J.-X.~Wang, Phys. Rev. {\bf D78} (2008) 074011. 
\bibitem{Lan11} J.P.~Lansberg, Phys. Lett. {\bf B695} (2011) 149-156. 
\bibitem{Lan09} J.P.~Lansberg, Eur. Phys. J. {\bf C61} (2009) 693.
\bibitem{Vog10} R.~Vogt, Phys. Rev. {\bf C81} (2010) 044903.
\bibitem{Kha11} V.~Khachatryan et al. (CMS Collaboration), Eur. Phys. J. {\bf C71} (2011) 1575. 
\bibitem{Aai11} R.~Aaij et al. (LHCb Collaboration), Eur. Phys. J. {\bf C71} (2011) 1645. 
\bibitem{Aad11} G.~Aad et al. (ATLAS Collaboration), Nucl. Phys. {\bf B850} (2011) 387.
\bibitem{Aam11} K.~Aamodt et al. (ALICE Collaboration), Phys. Lett. {\bf B704} (2011) 442.
\bibitem{Aam08} K.~Aamodt et al. (ALICE Collaboration), JINST {\bf 3} (2008) S08002.
\bibitem{Eta} $\eta=-\ln[\tan(\theta_{\rm lab}/2)]$, where $\theta_{\rm lab}$ is the polar angle in the laboratory frame.
\bibitem{Aam10} K.~Aamodt et al. (ALICE Collaboration), JINST {\bf 5} (2010) P03003.
\bibitem{Fac10} P.~Faccioli et al., Eur. Phys. J. {\bf C69} (2010) 657.
\bibitem{Gai82} J.E.~Gaiser, Ph.D. thesis, SLAC-R-255, 1982.
\bibitem{Abt09} I.~Abt et al, (HERA-B Collaboration), Eur. Phys. J. {\bf C60} (2009) 517.
\bibitem{Bos11} F.~Boss\`u et al., arXiv:1103.2394.
\bibitem{Aub03} B.~Aubert et al. (BaBar Collaboration), Phys. Rev. {\bf D67} (2003) 032002.
\bibitem{Aal09} T.~Aaltonen et al. (CDF Collaboration), Phys. Rev. {\bf D80} (2009) 031103; 
S.~Chatrchyan et al. (CMS Collaboration), arXiv:1111.1557.
\bibitem{Fac08} P.~Faccioli et al., JHEP {\bf 0810} (2008) 004; 
F.~Abe et al. (CDF Collaboration), Phys. Rev. Lett. {\bf 79} (1997) 578.
\bibitem{ButPC}  M.~Butensch\"on, proc. of 14$^{th}$ Int. Conf. on Hadron Spectroscopy", Munich, June 2011, arXiv:1109.1740; J.-X.~Wang, M.~Butensch\"on, priv. comm.
\end{thebibliography}
